\documentclass[pre,showpacs,preprintnumbers,amsmath,amssymb]{revtex4}

\usepackage[dvips]{graphicx}% Include figure files
\usepackage{dcolumn}% Align table columns on decimal point
\usepackage{bm}% bold math

\begin{document}
\title{Effects of  the low frequencies of   noise
on   On--Off intermittency 
}

\author{S\'ebastien Auma\^ \i tre$^1$, Kirone Mallick$^2$, 
Fran\c cois P\'etr\'elis$^1$
}%finauteur
%\auteurcourt{S. Auma\^ \i tre, K. Mallick, F. P\'etr\'elis}%finauteurcourt 
\affiliation{$^1$Laboratoire de Physique Statistique, CNRS UMR 8550,  Ecole Normale Sup\'erieure,
 24 rue Lhomond, 75005 Paris, France\\
$^2$Service de Physique Th\'eorique, Centre d'\'etudes de Saclay, 91191 
Gif--sur--Yvette Cedex, France} 
%\email{petrelis@lps.ens.fr}
\date{\today} 
\begin{abstract}
 A bifurcating system subject  to multiplicative noise can exhibit
on--off intermittency close to the instability threshold. For a
canonical system, we discuss   the dependence of this intermittency on
the Power Spectrum Density (PSD) of the noise.    Our study is based
on the calculation of the Probability Density Function (PDF)  of the
unstable variable.  We derive   analytical results for some particular
types of  noises and  interpret them  in the framework of on-off
intermittency. Besides,  we perform  a  cumulant expansion
\cite{VanKampen1} for a random  noise with arbitrary  power spectrum
density  and  show    that the intermittent regime is controlled  by
the ratio  between the departure from the threshold and the value of
the PSD  of the noise at  zero frequency.   Our  results are in
agreement with numerical simulations performed  with two  types  of
random perturbations: colored Gaussian noise and deterministic
fluctuations of a chaotic variable.   Extensions of this study to
another, more  complex, system  are  presented  and  the underlying
mechanisms are discussed.

\end{abstract}

%\authorrunning{S. Auma\^{\i}tre et al.}
%\titlerunning{Effects of the low frequencies of   noise on   on-off intermittency}

\maketitle
\email{petrelis@lps.ens.fr}
\section{Introduction} 
 
Most patterns  observed in nature are created by instabilities that
occur  in  an  uncontrolled noisy environment:  Convection in  the
atmospheric layers  and  in the mantle are subject  to  inhomogeneous
and fluctuating heat flux;  sand dunes are formed under winds  with
fluctuating directions and strengths. The  fluctuations  usually
affect   the control  parameters driving the  instabilities, such as
the Rayleigh number which is proportional  to the imposed temperature
gradient in natural convection. Thus,  these fluctuations  act
multiplicatively  on  the  unstable modes. In the same spirit, the
evolution of global quantities,  averaged under small turbulent
scales, can be represented  by a  nonlinear  equation with fluctuating
global transport coefficients   that  reflect  the small scales
complexity. For instance, it has been shown that the temporal
evolution of the total heat flux in rotating convection  can be
described by  a non--linear equation with a multiplicative  noise
\cite{Neufeld}. The dynamo instability   that  describes the growth of
the magnetic field of the stars and some planets because of   the motion
of conducting fluids  in their cores,   is usually  analyzed in
similar  terms:~the magnetic field  is expected to  grow  at  large
scale,  forced by a  turbulent flow. Here again,  the parameters
controlling the growth rate of the  field are fluctuating \cite{Sweet}.

Since the theoretical predictions of Stratonovich \cite{Strato}, and
 the experimental works of Kawaboto, Kabashima and Tsuchiya
 \cite{Kawakubo}, it is well known that a multiplicative  noise may
 modify an instability process.  These early investigations motivated
 numerous  studies on the effect of multiplicative noise on
 an instability threshold.
%By
%computing the PDF of the dynamical variables, 
It can be shown in many cases that  the noise induces a drift for the
instability threshold (see for instance
\cite{Schenzle,Horm,Lucke1,Kirone1,Sebetmoi}).  Besides, T. Yamada {\it et al}
\cite{Yamada} have shown that multiplicative noise can lead to a new
type of intermittency, called {\it On--Off Intermittency}, in which
quiet and  laminar  (off) phases  randomly follow  bursting (on)
phases.  This intermittency has been identified  in experiments in
various fields: electronics, electro-hydrodynamic convection in
nematics, gas discharge plasmas and spin-wave instabilities
\cite{Hammer}.

Most of the theoretical works considered only the effects of a
delta--correlated  Gaussian white noise or an Ornstein--Uhlenbeck
noise with an exponentially decaying correlation function (see for instance the discussion in \cite{Horm}). However,
with these  types of noises   that  have  at most one   characteristic
time scale, it is difficult to identify which part of the Power
Spectrum Density (PSD) of the random forcing really affects the
dynamics.  On the contrary,   the noise   in  natural environment  and also   in experimental situations   is far from being a white random
process. Therefore,  we believe that the influence of the  noise PSD  on
 an on-off intermittent dynamics  deserves to be investigated
 more precisely.

To motivate further reading of this article, we show in fig.\ref{fig1}
the temporal traces of an unstable variable subject  to two  different 
multiplicative noises.  Both noises have  the same standard deviation
but different power density spectra. More precisely, in fig.1a, the
PSD of the noise has a  higher value at zero frequency than in
fig.1b. It is clear  that the intermittent regime is  suppressed if
the low frequencies of the noise are reduced {\it even if} the
standard deviation of the noise is kept constant.   To understand this
fact, we study in Section  $2$ a  canonical system and  calculate the
PDF of the dynamical variable with different methods: exact results
for some special types of noises and a perturbative  expansion valid
for a small  noise amplitude. In Section  $3$, we compare the
predictions of this  expansion with  numerical simulations. We also
study the relation between the low frequencies of the noise PSD and
the statistics of the duration of the laminar phases in the
intermittent regime (Section  $4$).  In Section  $5$, we present
numerical simulations of a bifurcating system of second order in time.
We finally give a physical explanation for the relevance  of the noise
spectrum at zero frequency for  on-off intermittency (Section  $6$).

Some of the results of this article were published in our letter \cite{nous}. We give here details on the derivation of these results (Section $2$ and $3$). Besides new systems are investigated (Section $2.3$ and $5$) and a new aspect of the phenomenon is highlighted (Section $4$).
 
\section{Analytical predictions}

\subsection{Case of a Gaussian white noise}

We consider the simple  system proposed in \cite{Yamada} to describe
on--off  intermittency~:
 \begin{equation}
 \dot{X}=(a+\zeta(t)) X-X^3\,,
\label{eqbase}
\end{equation}
where $\zeta$ is a random process with zero mean.  This equation
describes the evolution of a variable $X$ with
instantaneous departure from onset  $a+\zeta(t)$ and cubic
nonlinearity.  Without noise ($\zeta=0$), equation~(\ref{eqbase}) has
the fixed points~: $X=0$  and $X=\pm \sqrt{a}$ for $a > 0$. The
former one is stable for negative $a$ and the latter are stable for
positive $a$.

Let $\zeta(t)$ be a Gaussian white noise with $\langle
\zeta(t)\zeta(t')\rangle_s= D \delta(t-t')$ where $\langle\rangle_s$
is the average on the realizations of the noise. The Langevin
equation~(\ref{eqbase}) is interpreted as a Stratonovich equation. 
The stationary
Probability Density Function of $X$ can be calculated from the
Fokker--Planck equation \cite{Schenzle} and is given by 
\begin{equation}
P(X)=C |X|^{\frac{2 a}{D}-1}e^{-\frac{X^2}{D}}\,,
\label{pdfgaussian}
\end{equation}
for $a > 0$;  $P(X)=\delta(X)$ if $a\leq 0$. Here, $C$ is a normalization constant.

Several features can be
noticed. For positive $a$, there are two different
behaviors. When $2 a > D$, the most probable values are $X_{mp}= \pm
\sqrt{a-D/2}\,$ but when $2 a \leq D$ the most probable values vanish
and $P(X)$ diverges as $X\rightarrow 0$. For a small departure from
threshold, i.e., $2a/D\ll 1$, $P(X)$ is dominated by a decreasing
power law over  a large range of $X$ and all moments of $X$ grow linearly
with $a$. Indeed,  equation~(\ref{pdfgaussian}) implies that
$\langle X^{2n} \rangle=D^n\Pi_{j=0}^{n-1}(a/D+j)$ which leads  to 
$\langle X^{2n} \rangle\simeq a D^{n-1}(n-1) ! $ 
 when $2a/D$ is small. 

As pointed out in \cite{Yamada}, the form of the PDF for small $X$ 
  is related to the  on--off intermittent
 character of the  variable $X$~: The occurrence of laminar phases
 are responsible for the divergence of the PDF at $X=0$.

\subsection{Expansion for a colored noise}

\begin{figure} %[!htb]
\includegraphics[width=8.6cm]{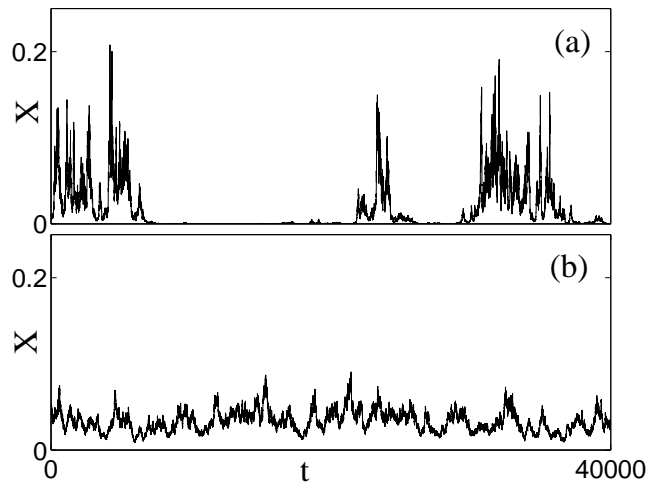}
\caption{ Temporal traces of the dynamical variable $X(t)$ solution of  equation
(\ref{eqbase}) with a noise of autocorrelation function  given by
(\ref{selfcorrbr}).  In both cases,   $a=0.00125$, $\alpha^2=0.005$.
In figure~(a), $\eta=\Omega=0.25$ i.e., $a/S=0.3927$; in   figure~(b),
$\eta=\Omega=2.5$ i.e.,  $a/S=3.9270$.
\label{fig1}}
\end{figure}

 White noise,  with   all frequencies having  the same weight, does
not allow to discriminate   which  frequencies play a   role in the
occurrence of on-off  intermittency.  However, as it clearly appears in
fig.\ref{fig1}, two non-white  noises with the same standard deviation
but different spectral densities  at  zero frequency, lead  to
dynamics  that are qualitatively  different. Indeed, if  the  value of
the noise PSD at  zero frequency is reduced,  the laminar phases
around zero, that  characterize  on--off intermittency, can even be
suppressed.

To analyze quantitatively this phenomenon, we apply the cumulant
 expansion to equation~(\ref{eqbase}).   The resulting equation for
 the PDF of $X$ is of the Fokker-Planck type and, in the case under
 study, is given by
\begin{equation}
\partial_t P= \partial_x\left(\left((1+\frac{S-M
}{a})X^3-(a+S)X\right)P\right) +\partial_{x^2}
\left(\left(S X^2+\frac{M-S}{a} X^4\right)P\right)\, .
\label{eqvankampen1}
\end{equation}
The derivation of this equation is presented in Appendix.  The  two
coefficients that appear in this effective Fokker-Planck equation
depend  on  the noise as follows
\begin{eqnarray}
S&=&\int_0^\infty \langle \zeta(0)\zeta(\tau)\rangle_s d\tau\nonumber
\, ,\\ M&=&\int_0^\infty \langle\zeta(0)\zeta(\tau)\rangle_s e^{-2 a
\tau} d\tau\, .
\label{defSM}
\end{eqnarray}
 The parameter $S$ is given by the integral of the autocorrelation
 function of the noise and is equal to  half of the PSD of the noise
 at zero frequency by virtue of  the Wiener--Khintchine theorem. The
 parameter $M$ is also related to the integral of the autocorrelation
 function but with a reduced weight of its long-time values.  The
 steady state  solution of equation~(\ref{eqvankampen1}) for the
 generic case $S\neq 0$ and $ S\neq M$ is given by
\begin{equation} 
P(X)=C |X|^{\frac{a}{S}-1} |1+\frac{(M-S)X^2}{S\, a}|^{-(1+\frac{a\,M}
{2\,S\,(M-S)})}  \, ,
\label{pdfcol}
\end{equation}
where $C$ is a normalization constant. Note that this expansion is
 valid when the product of the time correlation of the  noise with its
 amplitude is small \cite{VanKampen1,VanKampen2}.

The behavior of the PDF for small $X$ is a power law with exponent
 $a/S-1$. Consequently,  the criterion  for on-off intermittency, in
 the sense that the PDF of the variable diverges for small $X$, is
\begin{equation}
S>a\,.
\label{criteria}
\end{equation}
In other words, the variable is on-off intermittent when the value of
 the noise spectrum at zero frequency is  greater  than twice the
 departure from onset.

We also notice  from the power law form of the PDF that  all the moments  $<X^{2n}>$ grow linearly
with the departure from onset $a$, in the limit of small $a$.  As in the case of a Gaussian white
noise, this behavior  is  related to the form of the PDF in the vicinity
 of the unstable fixed point  and thus to the occurrence of on-off intermittency.

%\begin{figure}
%\begin{center}
%\epsfxsize=8.6cm
%\epsfbox{fig1ab.eps} 
%\caption{
%%\includegraphics[width=12cm]{fig1ab.eps}}
%%\caption{ 
%Temporal traces of the dynamical variable $X(t)$ solution of  equation
%(\ref{eqbase}) with a noise of autocorrelation function  given by
%(\ref{selfcorrbr}).  In both cases,   $a=0.00125$, $\alpha^2=0.005$.
%In figure~(a), $\eta=\Omega=0.25$ i.e., $a/S=0.3927$; in   figure~(b),
%$\eta=\Omega=2.5$ i.e.,  $a/S=3.9270$.}
%\label{fig1}
%\end{center}
%\end{figure}

\subsection{An exactly solvable case: the dichotomous Poisson  process}

It is also possible to calculate the PDF of $X$,  solution of equation
(\ref{eqbase}),   in the case where the noise is a dichotomous Poisson 
process. This problem was studied in  \cite{bruitdicho}. We sum it up
here and then discuss the consequences on the on-off intermittent
regime.

The noise has only  two possible values $\pm \Delta$ and during a time $dt$
 switches from one value to the other with a probability $\lambda dt$.
 We thus obtain 
\begin{equation}
<\zeta(t)\zeta(0)>_s=\Delta^2 e^{-2\lambda t}\,.
\label{corrDMN}
\end{equation}
Let $P_+(x,t)$ and  $P_{-}(x,t)$  be the probabilities for  the
 variable $X$ to attain  the value $x$ at time $t$ when  the noise is
 $\Delta$ and  $-\Delta$, respectively. These probabilities follow the
 equations
\begin{eqnarray}
\frac{\partial P_+}{\partial t}= -\frac{\partial}{\partial
x}\left(((a+\Delta) x-x^3)\,P_+\right)  -\lambda (P_+
-P_-)\,,\nonumber\\ \frac{\partial P_-}{\partial t}=
-\frac{\partial}{\partial x}\left(((a-\Delta) x-x^3)\,P_-\right)
-\lambda (P_- -P_+)\,.\label{eqnDMN}
 \end{eqnarray}

We consider  the case where $a$ is positive  so that 
 the fixed point  $X=0$ is  unstable. For
 intermittency to be possible, it is necessary that $\Delta >a$ so
 that the effective growth rate $a+\zeta(t)$ can be negative. In that
 case, the stationary PDF of $X$ is given by
\begin{eqnarray}
P(X)&=&P_{+}(X) +P_{-}(X)\nonumber\\
&=&C |X|^{\frac{2 a \lambda}{\Delta^2-a^2}-1} |X^2
-(a+\Delta)|^{\frac{\lambda}{2(a+\Delta)}-1}\nonumber\\ 
&&\times |X^2+\Delta -a|^{-\frac{\lambda}{2(\Delta-a)}-1}\,,
\label{eqPDFDMN}
\end{eqnarray}
where $C$ is a normalization constant.
 The PDF of $X$ diverges at small $X$ and therefore 
$X$ is on-off intermittent if 
\begin{equation}
\frac{2 a \lambda}{\Delta^2-a^2}<1\,.
\label{critDMN}
\end{equation}
From equation (\ref{corrDMN}) we calculate the parameter $S$~:
\begin{equation}
S=\int_0^{\infty}<\zeta(t)\zeta(0)>_s dt=\Delta^2/(2 \lambda)\,,
\end{equation}
and write the criteria for on-off intermittency as
\begin{equation}
S\ge \frac{a}{1-\frac{a^2}{\Delta^2}}\,.
\label{critbisDMN}
\end{equation}
 We emphasize  that this result is valid for any noise amplitude and
correlation time as long as $\Delta>a>0$. When 
 the product of $\Delta^2$ with  the  time
correlation $\lambda^{-1}$ of the noise is small, 
 we have $S \ll 1$ and  the criterion (\ref{criteria}) is
recovered. At higher noise amplitudes, we have an explicit expression
for the onset of on-off intermittency. Here again, if the parameter
$S$ is lowered and  the noise standard deviation $\Delta^2$ is fixed,
 on-off intermittency disappears.

\section{Numerical studies}

\subsection{Stochastic colored noise}

We verify  numerically the predicted expression for the PDF, 
 given in  equation~(\ref{pdfcol}). To wit, 
we use a colored noise with  two characteristic frequencies, $\Omega$
and $\eta$. This noise is generated from the following dynamics
\cite{Sawford} :
\begin{eqnarray}
\dot{A}&=&-4\pi\eta A-4\pi^2(\Omega^2+\eta^2)\zeta
+(4\pi)^{3/2}\sqrt{\eta(\Omega^2+\eta^2)/2}\alpha\xi\,,  \nonumber  \\ 
\dot{\zeta}&=&A\,,
\label{SDE}
\end{eqnarray}
where $\xi$  is a Gaussian white noise with $\langle \xi(t)\,\xi(t')\rangle_s=\delta(t-t')$. This
equation  leads to the following autocorrelation function
 \begin{equation}
  \langle \zeta(t)\,\zeta(t+\tau) \rangle_s =   \,\alpha^2 \exp(-2\pi \eta |\tau|)  \left(\cos(2\pi \Omega \tau) +
\frac{\eta}{\Omega} \sin(2\pi \Omega |\tau|)\right)\,,  \,\,\,\,
\,\,\, \label{selfcorrbr}
\end{equation}
where $\alpha^2$ is the noise variance  and $t_c=(2\pi \eta)^{-1}$ is
its correlation time.  In this  case, we obtain \cite{correc}
\begin{eqnarray}
S&=& \alpha^2\eta/\left[\pi(\eta^2+\Omega^2)\right]\,,\nonumber\\
M&=& \alpha^2(\eta+a/(2\, \pi))/\left[\pi\left((\eta+a/\pi)^2
+\Omega^2\right)\right]\,.
\end{eqnarray}
Therefore by  varying  $\eta$ and $\Omega$,  we can tune {\it independently}  $a/S$ and $\alpha(2\pi
\eta)^{-1}$.  The Gaussian white
noise is recovered in the limit $\eta\rightarrow {\infty}$ with
$\alpha^2/\eta=D$. The equations (\ref{eqbase}) and~(\ref{SDE}) are
solved numerically using a fourth-order Runge-Kutta scheme and an
Euler implicit method, respectively. Note from equation (\ref{eqbase})
that $X$ conserves its sign throughout its evolution. In the following, 
we consider only  positive initial values for  $X(t=0)$ without lack of
generality.

  In figure~\ref{fig1}, we  plot some temporal traces of $X$. 
   Both curves were obtained for the same
values of the noise variance $\alpha^2$  and   departure from
threshold $a$.  In fig.\ref{fig1}a, we have taken $S>a$; in
fig.\ref{fig1}b, the chosen value of $S$ is ten times smaller so that
the ratio $a/S$ becomes larger than unity. In the latter case,
intermittency is clearly suppressed, illustrating the fact that
 no intermittency occurs     when
  the PDF  $P(X)$ does not diverge at $X=0$.

\begin{figure}
%\begin{center}
%\epsfxsize=8.6cm \epsfbox{fig2.eps}
\includegraphics[width=8.6cm]{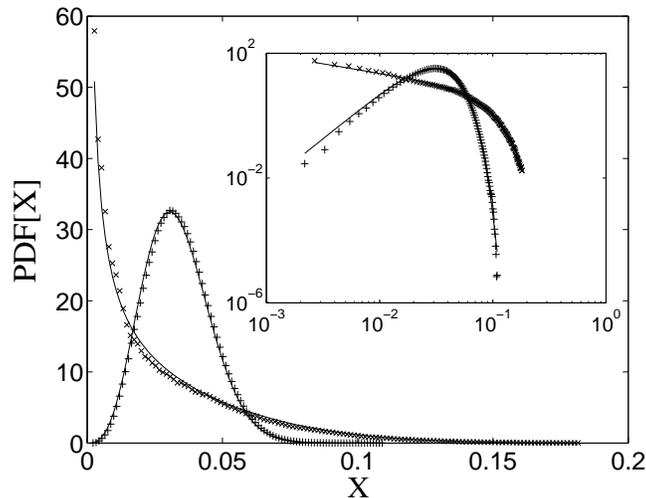}
\caption{PDF of the solutions of equation (\ref{eqbase}) for the 
colored  noise  given by~(\ref{selfcorrbr}) .  The symbols  $(\times)$
 and $(+)$  correspond, respectively,  to  the parameters  used  in
 figures~\ref{fig1}a and~\ref{fig1}b.  The full lines are the
 corresponding theoretical approximations   given by (\ref{pdfcol}).
 The inset emphasizes  the good agreement of the predicted power law
 in log--log axes.
\label{fig2}}
\end{figure}

 In figure~\ref{fig2}, we  show that the two  PDFs  corresponding 
 to the temporal traces of 
figures~\ref{fig1}a and \ref{fig1}b 
  are  very well described by equation~(\ref{pdfcol}). We 
  remark  that for   small values  of $X$,
 the PDF  behaves as  a negative power law when  $a/S<1$,
as expected in the intermittent regime.  

\begin{figure}
\includegraphics[width=8.6cm]{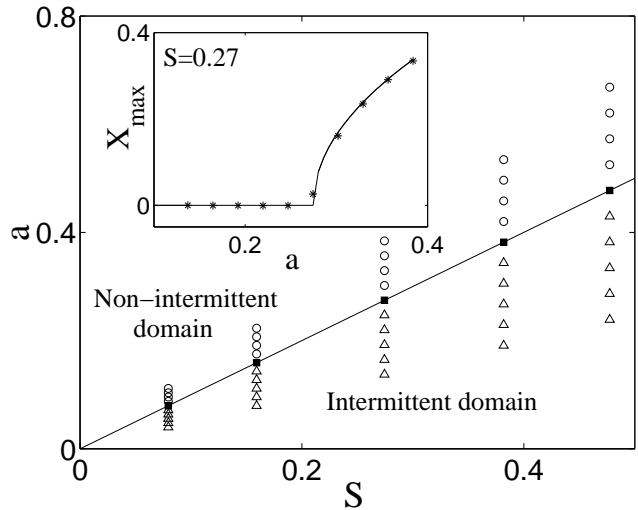}
\caption{ Boundary of the intermittent domain above the threshold $a\geq 0$
in the $(S,a)$--plane.  The open triangles $(\triangle)$ show the
 intermittent  domain where the most probable value $X_{max}$ of the
 PDF of $X$ is null.  The open circles   $(\circ)$ show the
 non--intermittent  domain where the most probable value $X_{max}$ is
 different from $0$.  The full squares $(\blacksquare)$ indicate the
 values of the parameters  for which  $X_{max}$ differs from $0$ for
 the first time in our simulations performed using increasing values of $a$.  The  straight line  $S=a$
 is the expected boundary predicted by   equation~(\ref{pdfcol}). The
 stars  $(\ast)$ in the  inset show  the most probable value
 $X_{max}$, as  obtained from  the numerical simulations  of
 equation~(\ref{eqbase}); the full line in the inset is the
 theoretical expression  $X_{max}=\sqrt{a(a-S)/(a+3(M-S))}~(a>S)$, which is
 derived from  equation~(\ref{pdfcol}).
\label{fig3}}
\end{figure}

 In figure~\ref{fig3},  the  intermittent domain and the
non--inter\-mittent domain  are delimited in the ($S,a$)--plane.
Intermittency disappears when the   most probable value, $X_{max}$,
becomes non-zero. The behavior of   $X_{max}$ as a function of $a$ for
$S=0.27$ is shown  in the inset of fig.\ref{fig3}.  For noises with different 
spectrum, we increase  $a$ and determine when on-off intermittency
disappears. We observe that the line $S=a$ does indeed separate the
two regimes.  Note that the expansion leading to equation
(\ref{criteria}) is valid when  $\alpha\tau_c\ll 1$, this condition is
fulfilled  in the simulations we present.

\subsection{Deterministic and chaotic  fluctuations as a noise}

  Up to now, the only  fluctuating parameters we have  considered  are
   stochastic  processes. However, it is  tempting to test the
   prediction of  equation~(\ref{pdfcol})   in the case of a
   deterministic but chaotic fluctuating parameter.  The noise is
   calculated from the chaotic solution of the Lorenz system
   \cite{Lorenz}. We thus solve
\begin{equation}
\dot{U}=-\sigma(U-Y)\,,\dot{Y}=r U-Y-U Z\,,\dot{Z}=U Y-b Z \, ,
\label{eqlorenz}
\end{equation}
and define $\zeta$ as
\begin{equation}
\zeta=\alpha\frac{(1-\mu) U_n +\mu\dot{U_n}}{c}\,,
\label{defzeta}
\end{equation}
where $c^2=<((1-\mu) U_n +\mu \dot{U_n})^2>$,
$U_n=\frac{U-<U>}{\sqrt{<(U-<U>)^2>}}$ and
$\dot{U_n}=\frac{\dot{U}-<\dot{U}>}{\sqrt{<(\dot{U}-<\dot{U}>)^2>}}$.
Averages are now understood as long time averages.  The role of $c$ is
to insure that $\alpha$ is the amplitude of the noise, i.e.,
$\sqrt{<\zeta^2>}=\alpha$.  The parameter $\mu$ is tuned  between zero
and one in order  to change the value of the spectrum at
zero-frequency. Indeed $\dot{U}$ being the derivative of $U$, its
power spectrum at  low frequencies is smaller than that of  $U$.
Increasing $\mu$ increases the magnitude of $\dot{U}$ and thus
reduces  the  spectrum of the  noise at low frequencies   (and
accordingly the value of $S$).

\begin{figure}
\includegraphics[width=9cm]{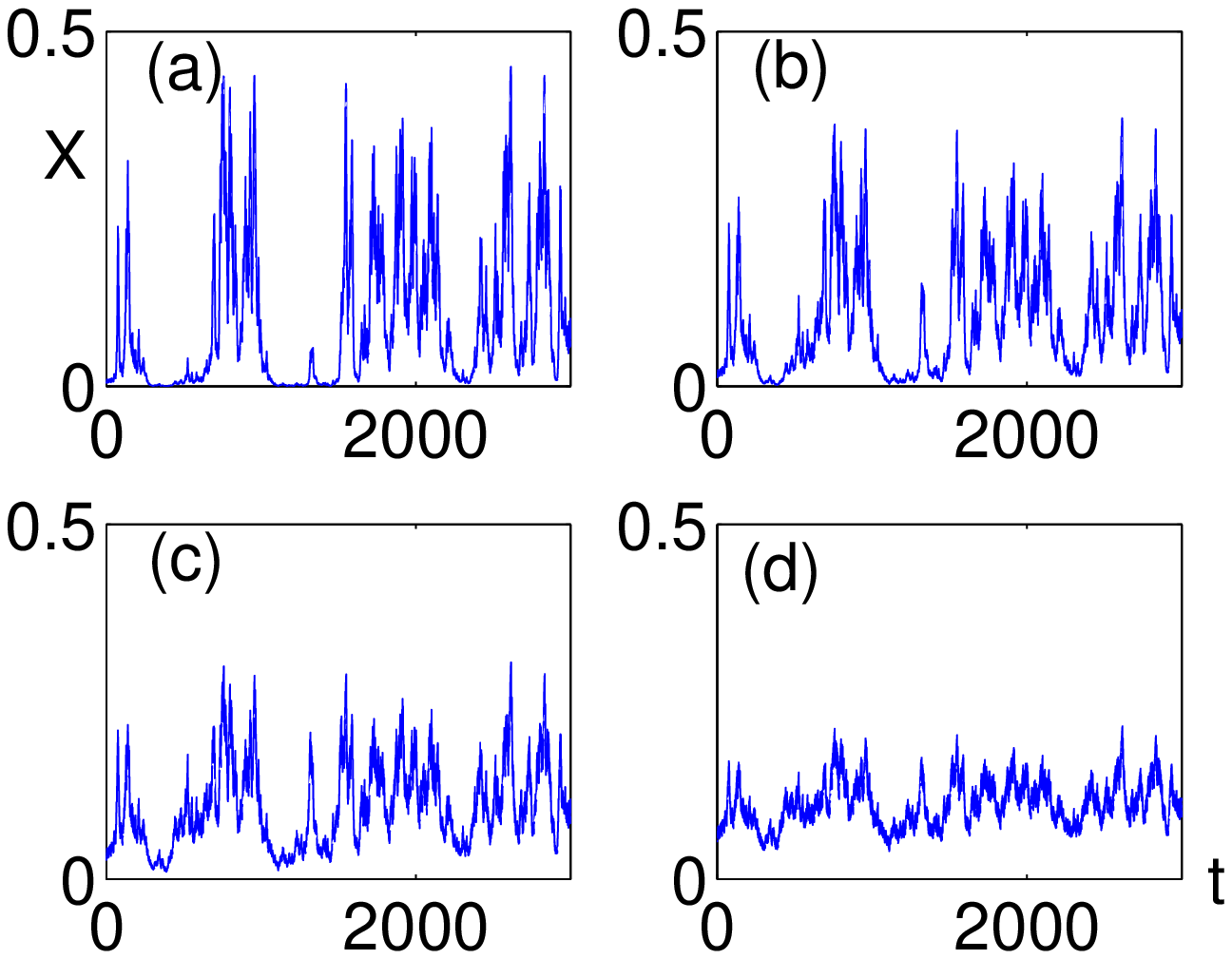}
\caption{Solutions of equation~(\ref{eqbase}) with $\zeta$ obtained from the
 Lorenz system (see figure \ref{fig7b}).  The departure from onset is
$a=0.01$, the  noise standard deviation is
$\sqrt{<\zeta^2>}=0.2$. (a): $\mu=0$, (b): $\mu=0.5$, (c): $\mu=0.65$,
(d): $\mu=0.8$.
%We have again  $\mu=0$ on the left and 
%  $\mu=0.8$  on the right. Note the difference in the horizontal scales $t$.}
\label{fig7}}
%\end{center}
%\end{figure}
%
%\begin{figure}
%\begin{center}
\includegraphics[width=9cm]{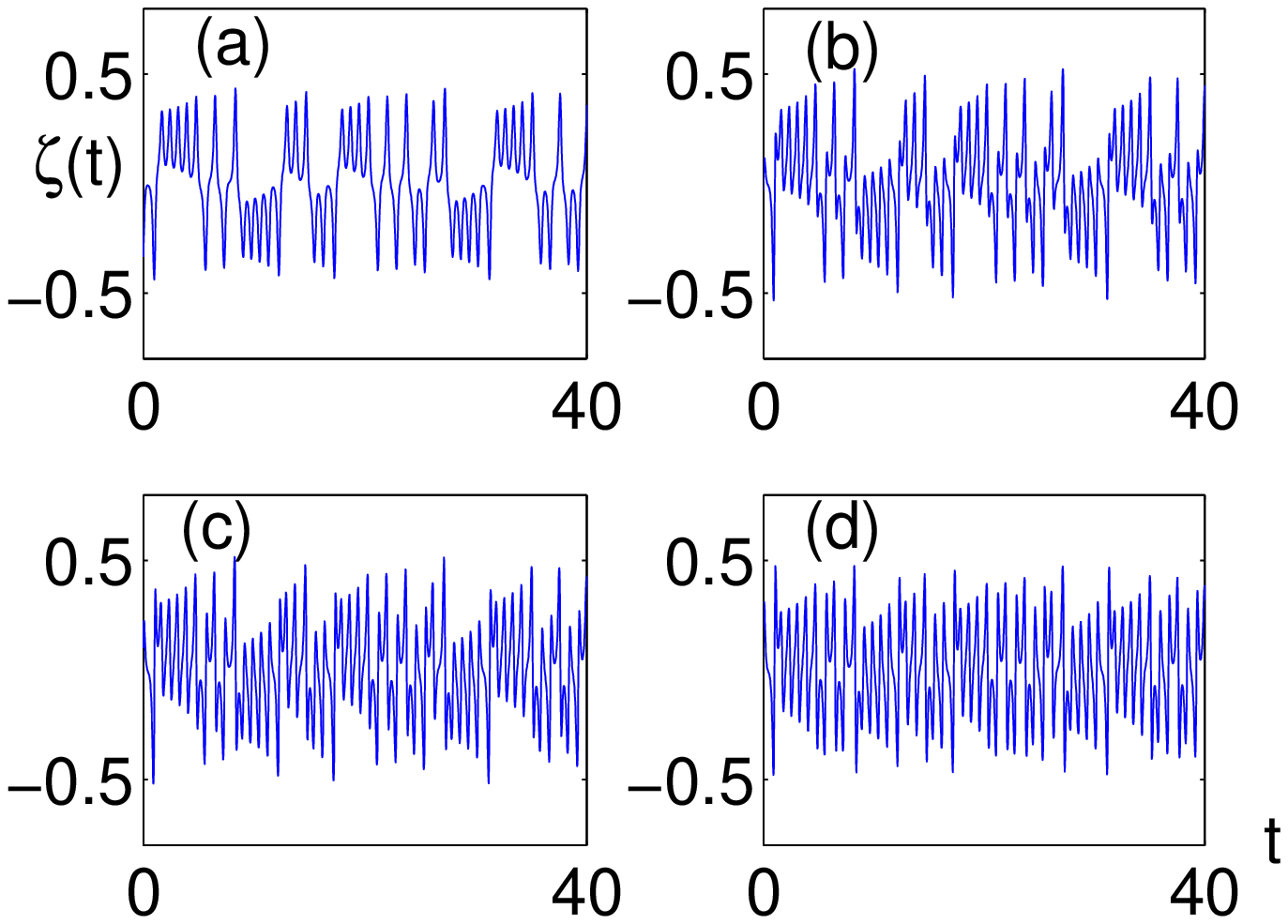}
\caption{Function $\zeta(t)$ obtained from the solution
 of the Lorenz system through equation (\ref{defzeta}).  (a): $\mu=0$,
(b): $\mu=0.5$, (c): $\mu=0.65$, (d): $\mu=0.8$.  Note the difference
in the horizontal scales $t$ with figure \ref{fig7}.
\label{fig7b}}
\end{figure}

\begin{figure}
\includegraphics[width=8.6cm]{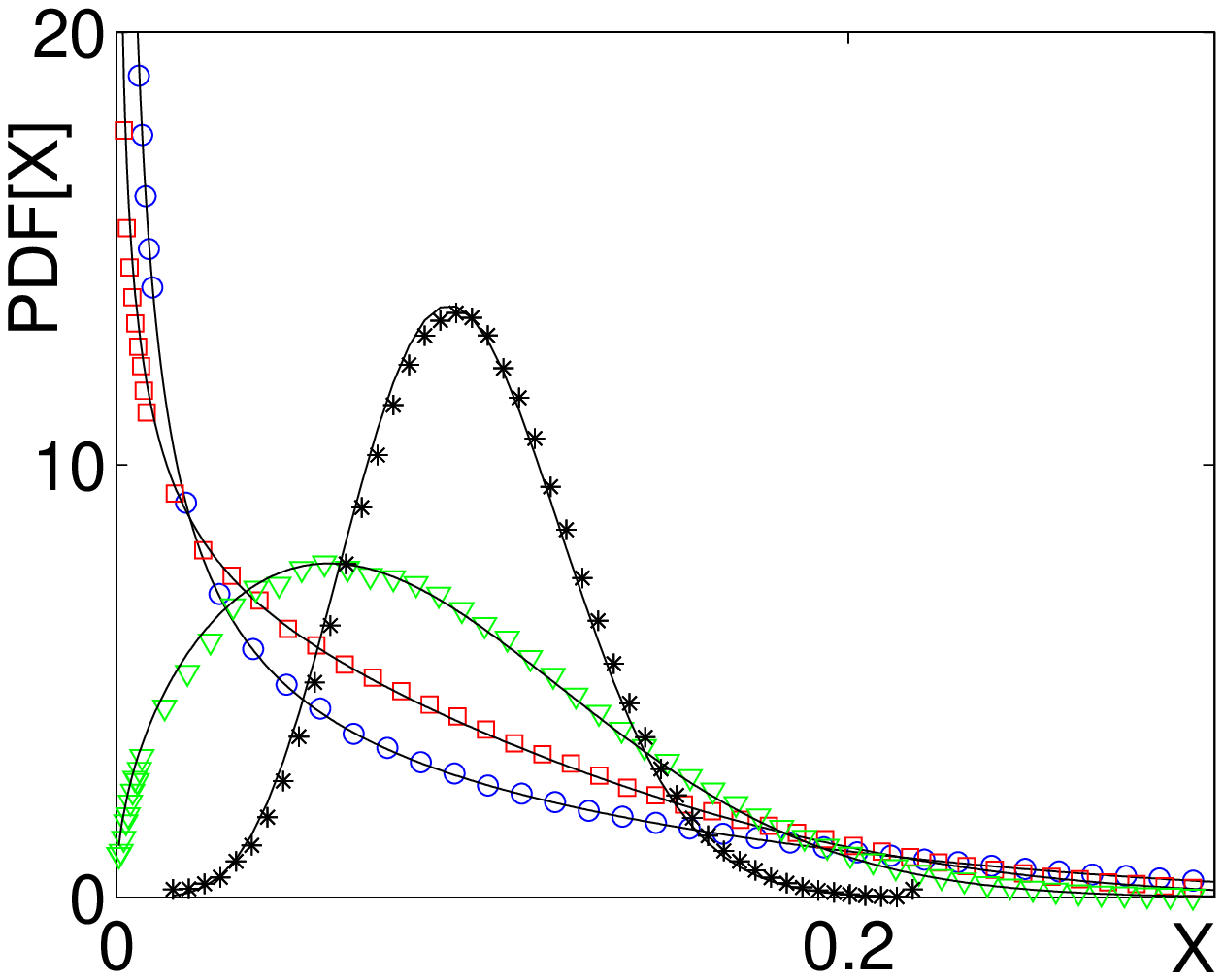} \includegraphics[width=8.6cm]{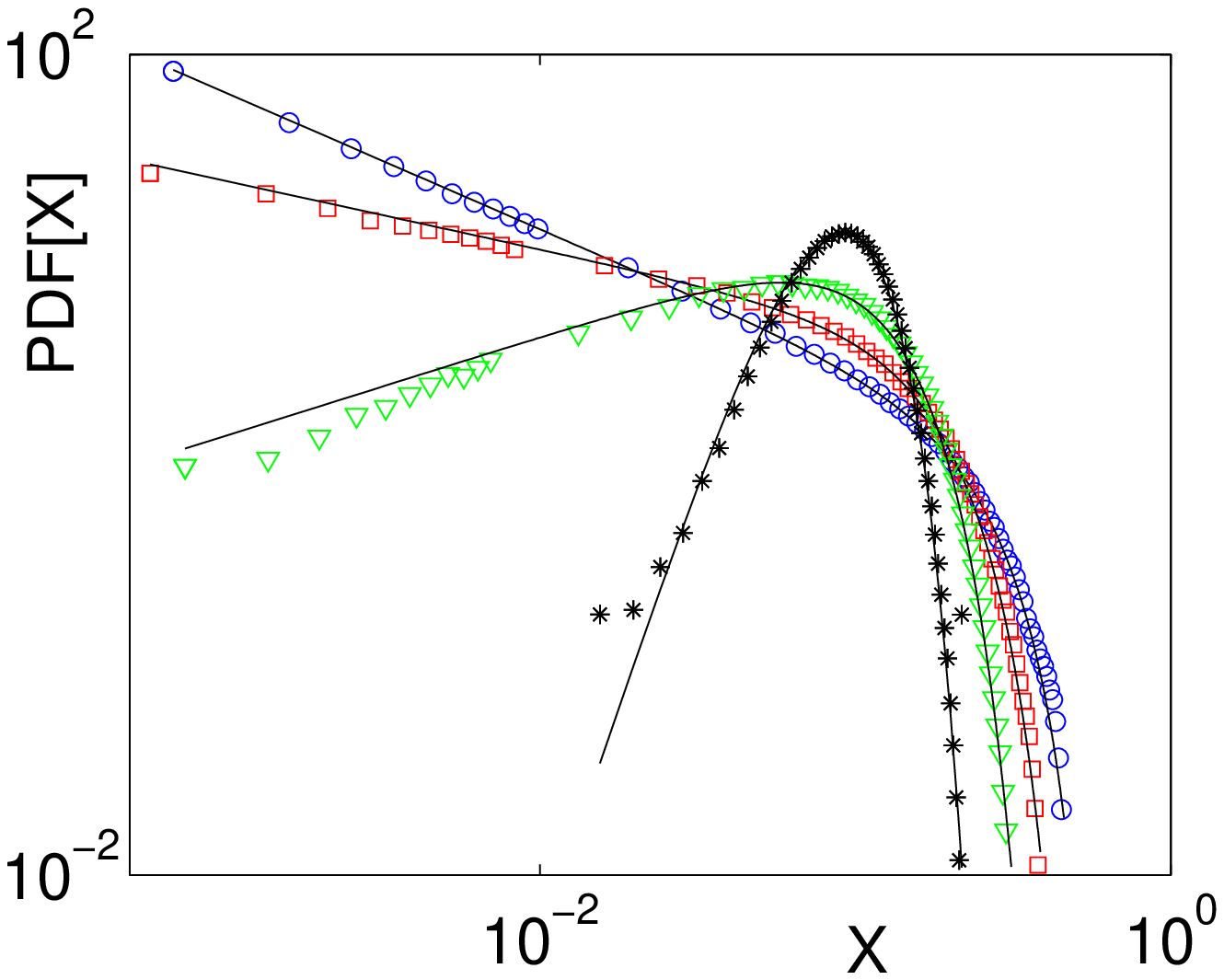}
\caption{Top figure: Probability density function of the solutions
 of equation (\ref{eqbase}) when $\zeta(t)$ is obtained  from the
Lorenz system through equation (\ref{defzeta}).  The  parameter values
are the same  as those of  figure~\ref{fig7}: $a=0.01$,
$\sqrt{<\zeta^2>}=0.2$.  The continuous lines are the theoretical
predictions  given by equation (\ref{pdfcol}).  The symbols represent
the numerical computation of the  PDF:  ($\circ$): $\mu=0$,
($\square$): $\mu=0.5$, ($\triangledown$): $\mu=0.65$, ($\ast$):
$\mu=0.8$.  Bottom figure: same results in loglog scale.
\label{fig8}}
\end{figure}

The equations (\ref{eqbase}, \ref{eqlorenz}) are solved with matlab
using the same methods as in section $3.1$. We choose $r=25$,
$\sigma=10$ and $b=8/3$. The solution of equation (\ref{eqlorenz}) is
then chaotic and we plot in figures~\ref{fig7} and~\ref{fig7b}  some
time series of $X$ and $\zeta$. On-off intermittency disappears when
$\mu$ increases and thus, accordingly,  $S$ decreases. This effect is
coherent with our former interpretation of the role of the  zero
frequency noise spectrum. Indeed we have $a/S=0.332$ for
fig. \ref{fig7b}a and $a/S=5.64$ for fig. \ref{fig7b}d. We also
 compute numerically  the PDF of $X$  and compare it with  the expression 
given by (\ref{pdfcol}).  The results are plotted in
fig. \ref{fig8}. There again, for small values  of the noise
amplitude, the agreement between the prediction and the numerical
results is very good.

\section{Statistics of  the durations of the laminar phases} 

The intermittent regime can also be  identified  by the statistics of
durations $\tau$ of the laminar  phases close to zero (see e.g.,
fig.\ref{fig1}a). We  discuss in this section  numerical results for
the durations  of the laminar  phases,  obtained  by using the random
process defined  in  equation (\ref{SDE}).

\begin{figure}
\includegraphics[width=8.6cm]{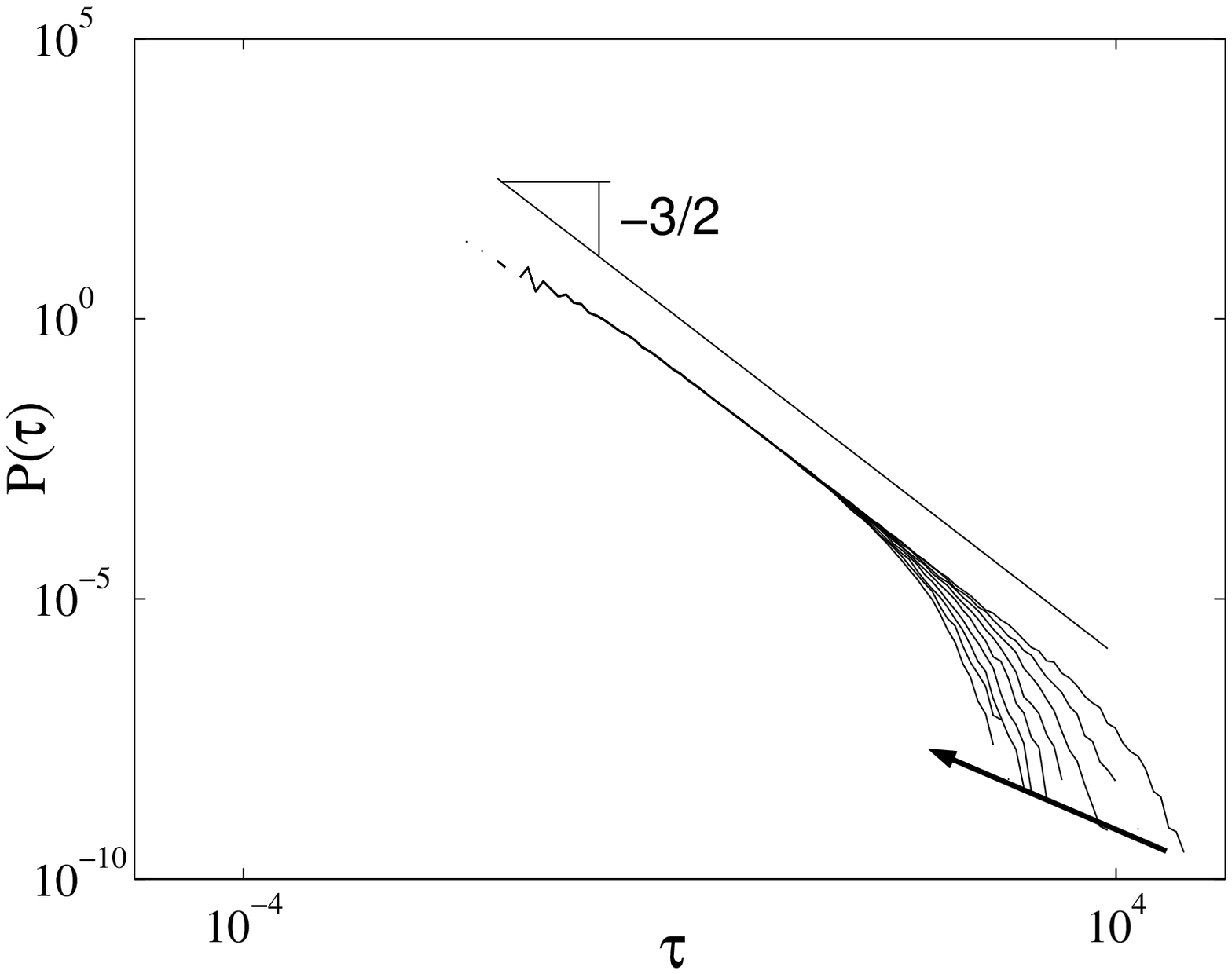}
\caption{PDF of the duration  $\tau$ of the laminar phases, for $S=0.159$
and $a$ varying   from 0.008 to 0.0796 $(=S/2)$. The arrow indicates the
curves   for increasing  values $a$.
\label{fig4}}
\end{figure}

 In the close vicinity of  the threshold, when  $a\rightarrow 0^+$, a
 power law with an exponent $-3/2$ is expected for the PDF of $\tau$
 \cite{Heagy}. This is in agreement with fig.\ref{fig4} where  we plot the PDF
 of $\tau$  for $S=0.159$ and for  various values of
 $a$. The threshold under which $X$ is considered to be in the laminar
 state is chosen arbitrarily  to be fifty times smaller than the noise
 intensity. However,  we have verified  that the PDF of $\tau$ does not
 depend strongly  on this choice if the
 threshold  remains   small enough compared to  the
 maximum of the bursts.

\begin{figure}
\includegraphics[width=8.6cm]{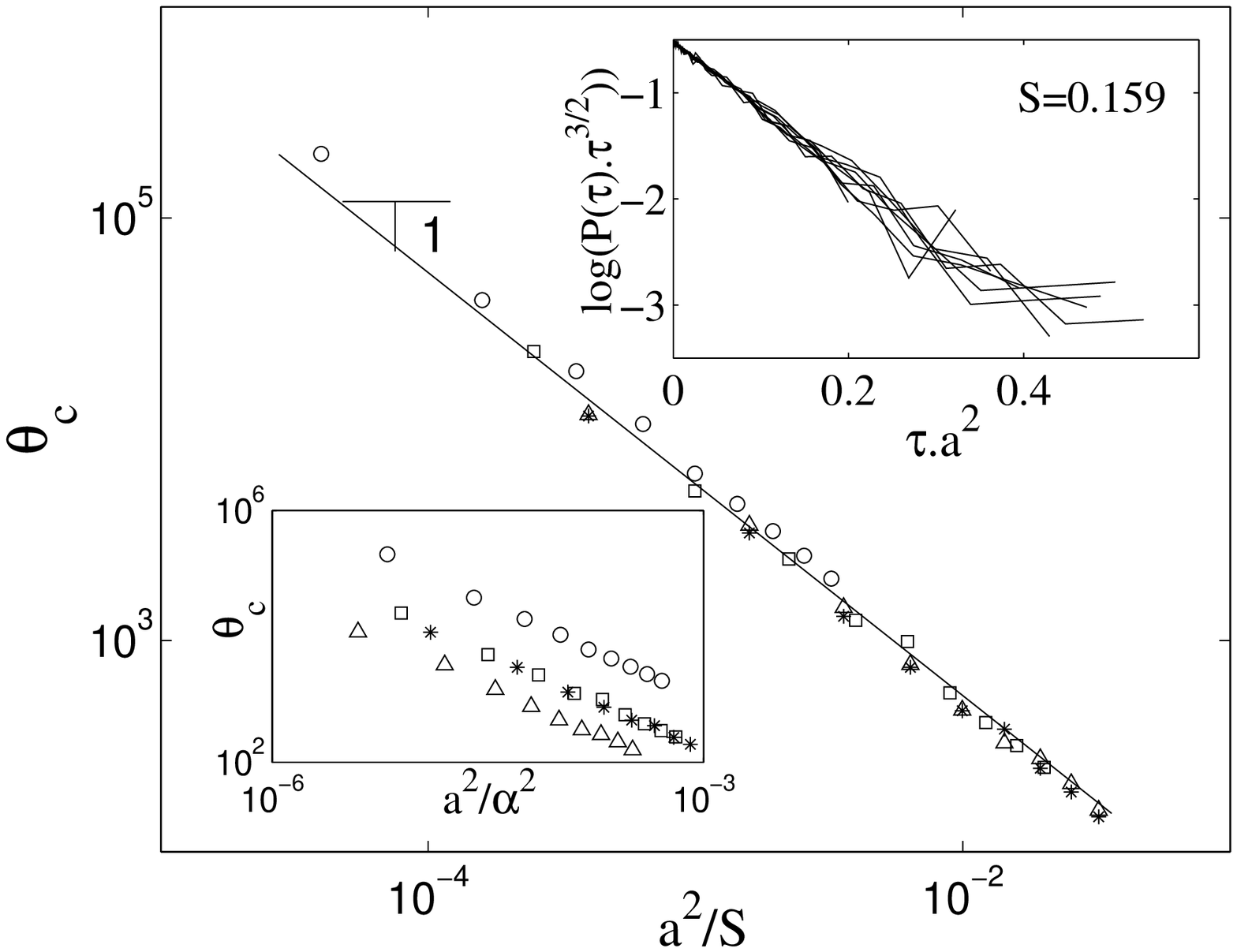} 
\caption{Characteristic time $\theta_c$ of the 
cut--off of the PDF of the laminar phase duration $\tau$,
plotted as a function of $a^2/S$. The values of the parameter are  $(\circ)$ : $S=0.0159,~\alpha^2=0.1$; 
$(\ast)$: $S=0.159,~\alpha^2=5.0$ ; $(\square)$ : $S=0.254,~\alpha^2=8$
and  $(\triangle)$: $S=0.159,~\alpha^2=8$. The upper--right inset
illustrates that the PDF of $\tau$ is well represented by
(\ref{eqPDFtau}). The lower--left inset shows that the  collapse is
 not obtained if $\theta_c$ is plotted as a function of $a^2/\alpha^2$.
\label{fig5}}
\end{figure}

We observe  that the cut--off takes  place at smaller
 values of $\tau$ when $a$ is increased.  
 More precisely fig.\ref{fig5} shows that the PDF of $\tau$ can be
fitted by
\begin{equation}
P(\tau)\propto \tau^{-3/2}.\exp(-\tau/\theta_c)\, ,
\label{eqPDFtau}
\end{equation}
where the characteristic time of the cut--off  $\theta_c$ is proportional to $S/a^2$. Indeed, the upper
right inset shows that  $\log\left(P(\tau)\right)+3/2\log(\tau)$ is
linear with $\tau$ in agreement with (\ref{eqPDFtau}). Moreover, the
central curve shows that all the characteristic times $\theta_c$
collapse on a single line if they are plotted as a function of
$a^2/S$.

This is in agreement with
 the exponential cut--off derived for white noise  in \cite{Heagy,Cenys}. In the white noise case, the PDF of $\tau$ follows equation (\ref{eqPDFtau}) with $\theta_c$ proportional to
 $D/a^{2}$ where D is the amplitude of the white noise. Our numerical studies  show that in the limit of small $S$ this prediction remains valid  for a non-white noise if $S$ is taken as the noise amplitude. 
Here again the noise power spectrum  at
zero frequency controls the value of $\theta_c$.  
As discussed in Part (VI.A), laminar phases occur when a random walk associated to the noise remains with the same sign for long durations. For small $S$ this property is controlled by the noise power spectrum at zero frequency. 

\section{Numerical simulations for a bifurcating system of 
   second order in time}

\begin{figure}
\includegraphics[width=8.6cm]{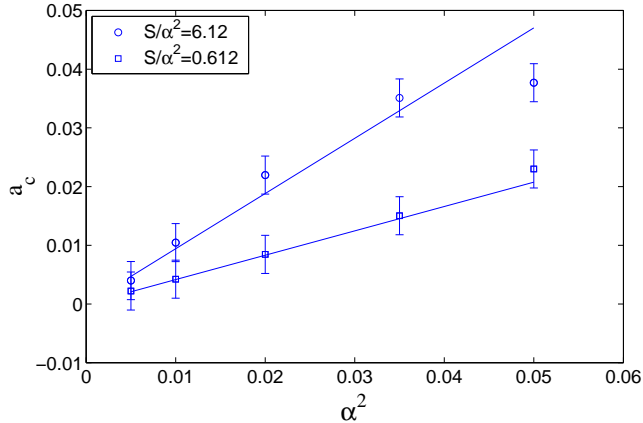}
\caption{Instability threshold of equation (\ref{FullDuff}) 
versus the noise intensity $\alpha^2$. The multiplicative noise  is
given by (\ref{selfcorrbr}), with ($\circ$): $\eta=0.05$, 
$\Omega=0.01$ and ($\Box$) : $\eta=0.5$,  $\Omega=0.1$.  Full lines
 represent the theoretical result~(\ref{Luckedrift}).
\label{drift}}
\end{figure}

The  Duffing oscillator and the effect of a multiplicative noise on
its dynamics have been widely studied.  Once the time is rescaled by
the viscosity, the Duffing oscillator perturbed by a  multiplicative
noise can be written as
\begin{equation}
\ddot{X}+\dot{X}=(a+\xi) X-X^3\,.
\label{FullDuff}
\end{equation}
L\"ucke and Schanck \cite{Lucke1} used an   expansion valid for a
small noise amplitude and  close to the deterministic threshold.  They
showed that a small amount of multiplicative noise  can stabilize the
state $X=0$ for positive $a$, whereas in the deterministic case,
$X=0$  is stable  only  for negative values of  $a$.  They calculated
the threshold shift induced by the  noise and found its expression as
a function of the noise Power Density Spectrum.  Their expansion leads
to the usual behavior  for the moments $\langle X^{2p}\rangle$ that
are proportional to the departure from onset raised to  the  power
$p$.  We emphasize that their  analysis is correct  only for noise
with a vanishing PSD at zero frequency \cite{Lucke2}.
 However, a recent study \cite{Kirone1} of the Duffing oscillator subject to 
 Gaussian white noise or  Ornstein-Uhlenbeck noise has predicted
 an intermittent behavior and a linear 
 scaling   of the moments $<X^{2p}>$  of the unstable variable 
  with the departure from onset. In order to clarify this 
 apparent contradiction between Refs.  \cite{Lucke1} and \cite{Kirone1}
 and to investigate   the effects of the low
frequency part of the noise spectrum on the Duffing oscillator,
 we study numerically  
  equation~(\ref{FullDuff}) with the colored  noise defined
 by (\ref{SDE}) for which the PSD is given by 
\begin{equation}
S(\nu)=\frac{ \alpha^2 \eta (\Omega^2+\eta^2)}
{\pi[(\Omega^2+\eta^2-\nu^2)^2+2\nu^2\eta^2]}\,.
\label{spectrecol}
\end{equation}
Contrary to the case studied in Sections 2-4,
 the onset of instability is shifted by the noise. 
We thus have to take into 
account the new threshold $a_c$. 
For small noise amplitudes, this threshold is given by \cite{Lucke1}
\begin{eqnarray}
a_c&=&\int_{-\infty}^{+\infty}\frac{S(\nu)}
{\nu^2+1/(2\pi)^2}d\nu\,,\\
&=&\frac{2 \alpha^2(1+4\pi\eta)}
{1+4\pi\eta+4\pi^2(\Omega^2+\eta^2)} \, .
\label{Luckedrift}
\end{eqnarray}  
 This theoretical result  agrees with the numerical data (figure
 \ref{drift}),  taking into account the uncertainty  in  the numerical
 determination of the threshold.

\begin{figure}
\includegraphics[width=8.6cm]{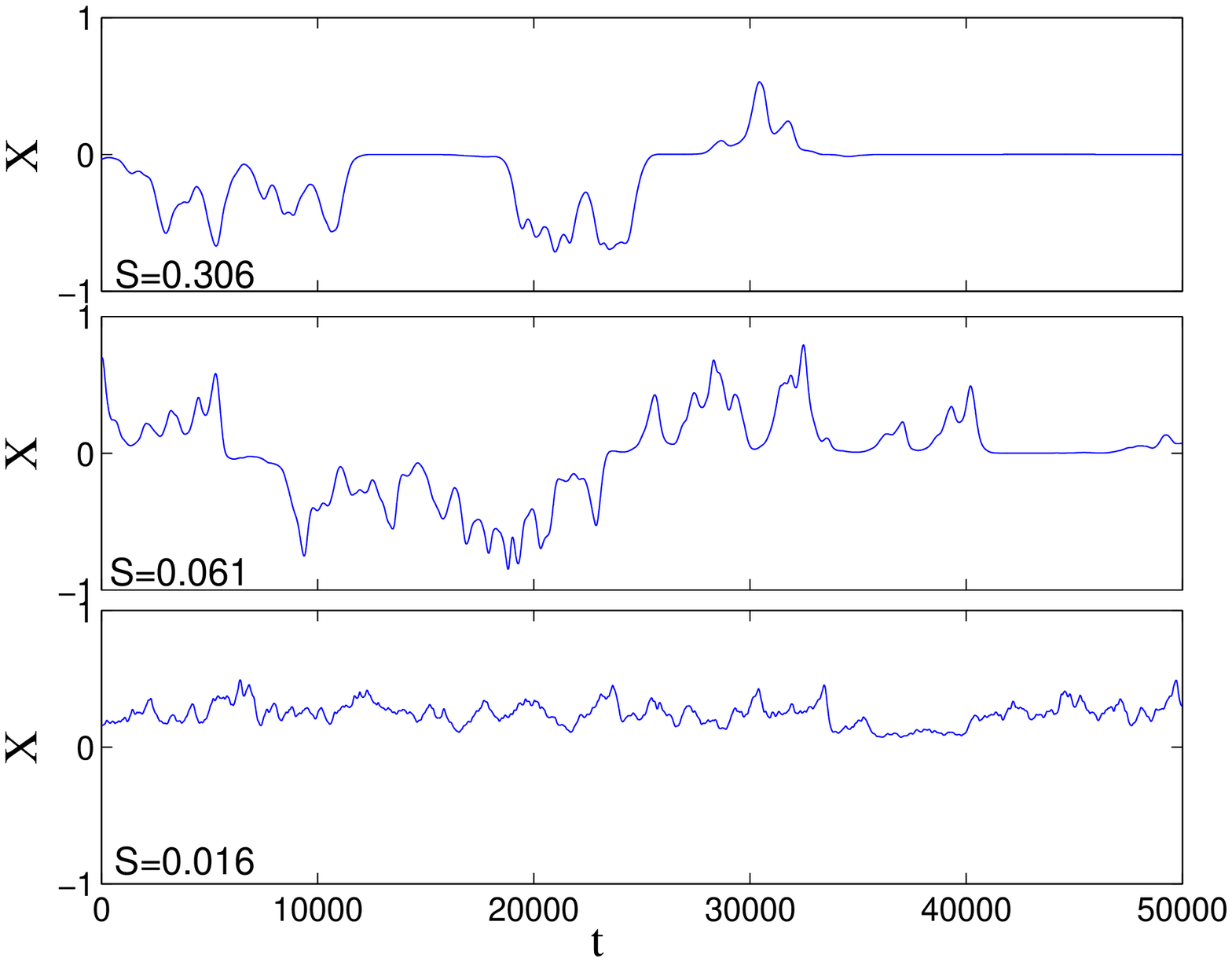}
\caption{
Temporal traces of the dynamical variable $X(t)$,  solution of
equation (\ref{FullDuff}). In all cases  $a-a_c=0.0754$,
$\alpha^2=0.05$ and the  autocorrelation function of the noise is
given by (\ref{selfcorrbr}), but $S$ is decreased from top to
bottom.
\label{smallX}}
\end{figure}

Figure \ref{smallX} shows  the temporal trace of $X(t)$  above
onset. It emphasizes  the fact that  $S$ is still the pertinent parameter
controlling the intermittent regime for small noise, i.e., for 
$\alpha^2<<1$. The same behavior is  observed for the temporal
trace of the other dynamical variable $\dot{X}(t)$.

\begin{figure}
\includegraphics[width=8.6cm]{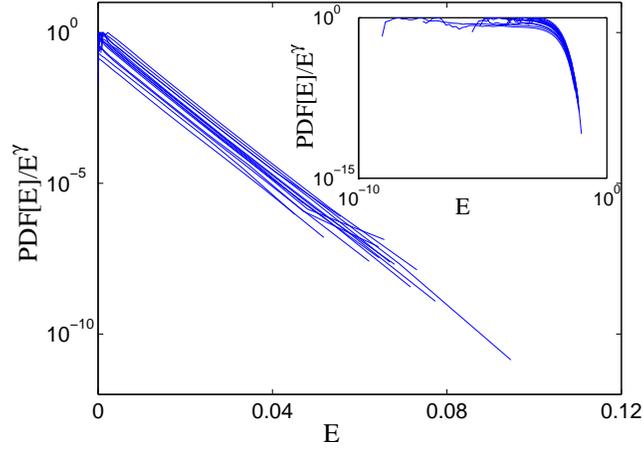} 
\caption{Probability Density Function of $E=X^2+\dot{X}^2$ divided by
$E^{\gamma}$ with $\gamma=\frac{a-a_c}{2S}-1$ 
 and for $\frac{a-a_c}{2S}$ 
going from 0.6 to 5.0. The inset in log--log plot underlines the plateau 
where the power law dominates the PDF.
\label{PDFE0005}}
\end{figure}

 Besides,  figure \ref{PDFE0005} shows that  the statistical behavior
 of the variable $E=X^2+\dot{X}^2$  is similar  to  that of the
 variable $X^2$ in the first order system studied in Sections 2-4.
 Indeed, the PDFs of $E$  divided by $E^{\gamma}$ with
 $\gamma=\frac{a-a_c}{2S}-1$ collapse on a single exponential  for
 various values of  $a$.  Notice that the departure from the onset in
 the presence of noise must be taken into account.  Therefore,  when
 the amplitude of the noise is small, the PDF of the energy is
 controlled by the ratio  between the departure from onset (in the
 presence of noise) and the value of the noise spectrum  at zero
 frequency.  When the amplitude of the noise is large, the PDF of the
 energy does not take the form suggested in
 fig. \ref{PDFE0005}. However,    even if the noise amplitude is
 large, on-off intermittency disappears  when the value of $S$ is lowered.

To conclude this part, we point out that the failure of the
perturbative expansion 
 \cite{Lucke2} and the linear scaling  of the moments as a function
of the departure from onset \cite{Kirone1} are both a consequence of
 on-off intermittency  that occurs  when the noise 
 is  sufficiently large   at low  frequencies.

\section{Physical interpretations and summary}

\subsection{Role of the low frequencies of the noise}

In the different systems we have studied, on-off intermittency
  is controlled by the  zero frequency component of the noise.
Our interpretation of the phenomenon is the following. On-off
intermittency occurs because of a competition between the noise and a
systematic drift due to the departure from onset. More precisely, as
pointed out in \cite{Yamada} for the case of
 equation (\ref{eqbase}), when $X$ is close to the unstable
manifold $X=0$, the evolution of $Y=\log{X}$ is given by
$\dot{Y}=a+\zeta(t)$. For positive $a$, $\dot{Y}$ has a positive
average but events in which $Y$ has a decreasing behavior are possible
provided that $I=\int_0^T \zeta(t) dt /T$ remains smaller than $-a$
 over  a  long duration. In the long time limit, the main contribution to
the integral $I$ is due to the zero frequency component of the
noise. If this component is reduced then occurrences of the inequality
$I\le-a$ become less and less probable and intermittency tends to be
suppressed.

\subsection{Linearity of the moments}

 We now want to explain why, close to the onset of instability, all the
 moments vary  linearly  with $a$, the departure  from onset. One can
 say that this is a direct consequence of the form of the PDFs that are
 power laws with exponents close to $-1$, the  difference from $-1$
 being proportional to $a$ (see  equations (\ref{pdfgaussian}),
 (\ref{pdfcol}), (\ref{eqPDFDMN})).  However,  we look here for an explanation
 based on the dynamical properties of the trajectories $X(t)$.

\begin{figure}
\includegraphics[width=8.6cm]{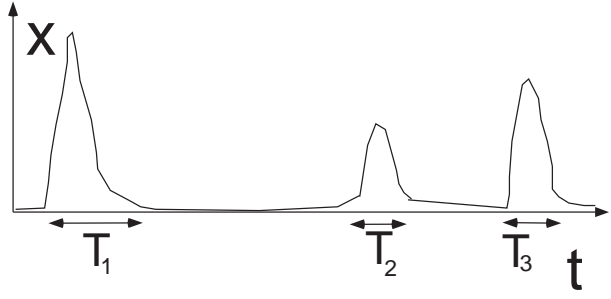} 
\caption{Sketch of the intermittent regime with $T_i$  the duration of the
$i^{th}$ burst.
\label{figtyptraj}}
\end{figure}

In the small $a$ limit, the variable $X$  spends  long 
 durations  in the
off-phase and, from time to time,  it takes non-zero values.
  A typical trajectory  is sketched  in 
 fig.~\ref{figtyptraj}. Let $T_i$ be
the duration of the $i$-th  on-phase  and $T_e=T_1+T_2+..$ be 
the total  time spent in the on-phases during the measurement time $T$.

\begin{figure}
\includegraphics[width=8.6cm]{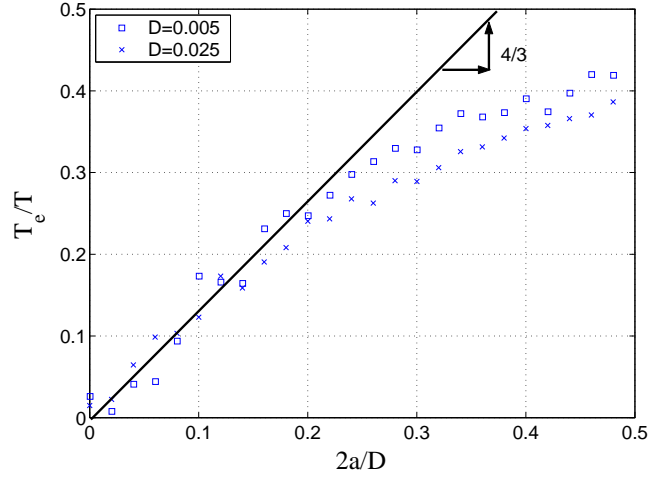} 
\caption{Total duration of the on--phase, $T_e$, normalized by the total 
duration of the measurement, $T$, as a function of  the departure from the 
threshold, $a$, given for two values of the white noise intensity 
$\langle\xi(t)\xi(t')\rangle=D\delta(t-t')$.
\label{figTeoverT}}
\end{figure}

 During the on-phases the evolution of $X$ can be described  approximately
  by a random walk with a drift 
 in terms of the variable  $Y=\log{X}$; besides,  the effect of
 nonlinearities can be modeled by a wall that prevents $Y$ from
 reaching too high values.
 Let us call  $C_n$ the averaged  value taken by $X^n$ during an
on-phase.  
Using the fact that the off-phases  have a negligible contribution
  to $\langle X^n\rangle$, we can  write approximatively
 for a large measurement time $T$
%Using the fact that there is almost no contribution of the
%off-phases to any moment, one can write approximately for large
%measurement time $T$
\begin{equation}
\langle X^n \rangle=\frac{1}{T}\int_0^T X^n(t') dt'\simeq
\frac{T_e}{T} C_n\,.
\label{eqmomapp}
\end{equation}
For large $T$, $T_e/T$ is the product of the averaged duration of an
on-phase with the averaged  frequency of occurrence of an on-phase. Using
the aforementioned analogy with a biased random walk limited by a
wall, we conclude that the averaged  duration of an on-phase is finite
when the drift $a$ tends to zero. Moreover, the averaged frequency of
occurrence of an on-phase is proportional to 
 $a$ and  therefore  $T_e/T$ is also
proportional to $a$. This scaling law is tested numerically  for  
  equation~(\ref{eqbase})  with Gaussian  white noise.
  We plot  in fig. \ref{figTeoverT} the quantity   $T_e/T$ as a function of $a$~: 
 The relation is  linear  when  $a/D$ is small.

Finally,  $C_n$ the averaged value of the $n$-th moment of $X$  during the
 on-phase can be calculated in the case of a Gaussian white noise using  eq.~(\ref{pdfcol});  $C_n$  tends
 to a  non-zero  constant when $a$ tends to zero. This fact  can be
 understood   using the analogy with the biased random walk limited by
 a wall: the typical trajectories restricted between the onset of the
 on-phase and the wall do not depend on $a$ for vanishing $a$.

 To summarize, when $a$ is very small, the system enters  on-phases with a
 frequency linear with $a$. However,  the duration of these
 on-phases and the values reached by the system during these phases do
 not depend on $a$.  Therefore,  using  eq. (\ref{eqmomapp}) and the
 above  discussion, we conclude that
\begin{equation}
\langle X^n \rangle\propto a \, ,
\label{eqmomappbis}
\end{equation}
 i.e.,  all  the moments are linear with the departure from onset.

\subsection{Summary}
We have studied different bifurcating systems subject to
multiplicative noise. For a system of first order in time  and for a
small value of the product of the noise amplitude with its correlation
time, an expansion showed that  on-off intermittency occurs if the
noise spectrum at zero frequency is greater  than twice the departure
from onset.  This prediction is in agreement with numerical
simulations that use colored random processes or chaotic fluctuations
as noises. In the same limit we have shown that the statistics of the
durations of the laminar phases are also controlled by the departure
from onset and the noise spectrum at zero frequency.   Even at finite
amplitude of the noise, we have verified numerically that
intermittency disappears when the low
frequencies of the noise are filtered out. 
This result is also derived analytically for a Gaussian white noise and 
for another particular kind of
noise, the dichotomous Markovian process.  For a system of second
order in time, we have numerically studied the behavior of the
unstable variable and showed that for small noise amplitudes, the PDF
of the energy scales as a power law with  exponent  controlled by the
noise spectrum at zero frequency and the departure from the onset.  
  Here again,  by lowering the noise spectrum at
zero frequency, the on-off intermittency is reduced
 and can be suppressed.  Finally,  we have
given  some physical explanations for  the effect of the noise
spectrum at zero frequency on  on-off intermittency and for  the
behavior of all  the moments of an on-off intermittent variable that
are linear with the departure from onset.

This work has benefited from fruitful discussions
 with C. Van den Broeck, P. Marcq, N. Leprovost and S. Fauve.

\section*{Appendix~: derivation of the cumulant expansion for a 
dynamical system of first order in time } 

If we consider one realization of the noise $\zeta(t)$ as a single
time dependent forcing, then for a given initial condition  $X(t=0)$,
equation (\ref{eqbase}) describes a single trajectory. In other words,
for a given realization of the noise,   the number of  trajectories  in
phase--space is conserved.  A  continuity equation for  the density of
trajectories  in the phase--space $\rho_\zeta(X,t)$ can therefore be
written \cite{VanKampen1} as follows~:
\begin{eqnarray}
\partial_t\rho_\zeta(X,t)&=&-\partial_{X}\left[\dot{X}\rho_\zeta(X,t)
\right]  \nonumber   \\  
&=&-\partial_{X}\left[(aX-X^3)\rho_\zeta(X,t)\right]  \nonumber  \\ 
&&-\alpha \partial_{X}\left[F_\zeta(X,t)\rho_\zeta(X,t)\right]
  \, , \label{contApp}
\end{eqnarray}
where $\alpha$ is the standard  deviation of the noise and
$$F_\zeta(X,t) = \zeta(t)X(t)/\alpha .$$ The PDF of $X(t)$ is just the
average of $\rho_\zeta(X,t)$ over  all the realizations  of the
noise. Therefore,  by averaging  eq.~(\ref{contApp}),  an 
 evolution  equation for $P(X,t)$ can be derived.
 Some  approximations are however
necessary to obtain an equation which is closed 
 with respect to  $P(X,t)$. In  \cite{VanKampen1}
(pp~210),  Van Kampen expands  equation~($25$)  in powers  of the parameter 
$\epsilon=\alpha^2\tau_c$ where  $\tau_c$  is the correlation time of the
noise.  Assuming  that 
 $\epsilon<<1$ and knowing that  $\langle
F_\zeta(X,t)\cdot F_\zeta(X',t')\rangle\sim 0$ for $|t-t'|>\tau_c$,
 the following equation for $P(X,t)$ is derived~:

\begin{eqnarray}
&\partial_t P(X,t)=-\partial_{X}\left[-(aX-X^3) P(X,t)\right]\\\nonumber
&-\alpha^2\partial_{X}\left\{\left[\int_0^\infty \langle 
\frac{\partial F_\zeta(X,t)}{\partial X }
F_\zeta(X^{-\tau},t-\tau)\rangle \left|\frac{dX}{dX^{-\tau}}\right|
\right] P(X,t)\right\}\\\nonumber
&-\alpha^2\partial^2_{XX}\left\{\left[\int_0^\infty \langle 
F_\zeta(X,t)
F_\zeta(X^{-\tau},t-\tau)\rangle \left|\frac{dX}{dX^{-\tau}}\right|
\right] P(X,t)\right\}\,,
\label{VK19_8App}
\end{eqnarray}
where $X^{-\tau}=X_o(t-\tau)$ is the {\it deterministic backward position}, 
i.e., $X^{-\tau}$ represents the value of  the variable $X$ at time 
  $(t-\tau)$ such that $X$  would  
  evolve upto   $X_o(t)$   during the duration  $\tau$ if there were no noise. 
  The quantity $\left|\frac{dX}{dX^{-\tau}}\right|$ is
  the Jacobian of $X$ with respect to $X^{-\tau}$. Equation 
($26$)  is a second order expansion   in power  of 
the small parameter $\epsilon=\alpha^2\tau_c$
and  is therefore   valid as long as  $\epsilon<<1$. 

  For  equation~(\ref{eqbase}), all the
 quantities such as  $X^{-\tau}$ and the Jacobian can 
be explicitely calculated. By solving (\ref{eqbase}) 
with $\zeta(t)=0$, we find that  
\begin{eqnarray}
X&=&\pm\left\{\left[(a(X^{-\tau})^{-2}-1)\exp(-2a\tau)
+1\right]/a\right\}^{-1/2}\nonumber  \, ,\\ 
X^{-\tau}&=&\pm\left\{\left[(aX^{-2}-1)\exp(2a\tau)
+1\right]/a\right\}^{-1/2}\nonumber \, ,\\  \nonumber
\left|\frac{dX}{dX^{-\tau}}\right|&=&(X/X^{-\tau})^3 \, .
\label{Soleqbase}
\end{eqnarray}
  Finally, equation (\ref{eqvankampen1})
 is obtained   by substituting the  expressions for 
$\left|\frac{dX}{dX^{-\tau}}\right|$ 
and $X^{-\tau}$ in equation~($26$) 
 and writing  $F_\zeta(X,t)=\zeta(t)X(t)/\alpha$. 

\vfill 
\begin{thebibliography}{}

\bibitem{VanKampen1} 
N.G. van Kampen, 
Physics Reports {\bf 24} 171 (1976).

\bibitem{Neufeld} 
M. Neufeld, R. Friedrich, Phys. Rev. E, {\bf51}, 2033 (1995).

\bibitem{Sweet} 
D. Sweet, E.Ott, J.M. Finn, T.M. Antonsen Jr, D.P. Lathrop,
Phys. Rev. E, {\bf63}, 066211 (2001), D. Sweet, E.Ott, T.M. Antonsen Jr, 
D.P. Lathrop, J.M. Finn, Phys. Plasma, {\bf8}, 1944 (2001). S. Fauve and F. P\'etr\'elis, ``The dynamo effect'', pp. 1-66,
``Peyresq Lectures on Nonlinear Phenomena, Vol. II'', Ed. J-A Sepulchre,
World Scientific (Singapour, 2003).        

\bibitem{Strato}
R.L Stratonovich, {\it Topics in the Theory of Random Noise} (Gordon and 
Breach, New--York, 1963).

\bibitem{Kawakubo}
T. Kawakubo, S. Kabashima, Y. Tsuchiya, Prog. Theor. 
Phys. supp., {\bf64}, (1978).

\bibitem{Schenzle} 
A. Schenzle, H. Brand, Phys. Rev. A, {\bf20}, 1628 (1979),
R. Graham, A. Schenzle, Phys. Rev. A, {\bf26}, 1676 (1982).

\bibitem{Horm} W. Horsthemke, R. Lefever, {\it Noise-Induced Transitions} (Springer-Verlag, 1984).
   
\bibitem{Yamada}
T. Yamada, H. Fujisaka, Prog. Theor. Phys., {\bf76}, 582 (1986),
H. Fujisaka, H. Ishii, M. Inoue, T. Yamada, Prog. Theor. Phys., {\bf76}, 
1198 (1986), N. Platt, E. A. Spiegel and C. Tresser, Phys. Rev. Lett. {\bf 70} (3) 279-282 (1993).


\bibitem{Hammer} 
P. W. Hammer, N. Platt, S. M. Hammel, J. F. Heagy and B. D. Lee, Phys. Rev. Lett. {\bf 73}, 1095 (1994), 
T. John, R. Stannarius and U. Behn, Phys. Rev. Lett. {\bf 83}, 749 (1999),
D. L. Feng, C. X. Yu, J. L. Xie and W. X. Ding, Phys. Rev. E {\bf 58}, 3678
(1998),
F. R\"odelspreger, A. Cenys and H. Benner, Phys. Rev. Lett. {\bf 75}, 2594
 (1995).

\bibitem{nous} S. Auma\^{\i}tre, F. P\'etr\'elis and K. Mallick, Phys. Rev. Lett. {\bf 95}, 064101 (2005).

    
\bibitem{Lucke1} 
M. L\"ucke, F. Schanck,  Phys. Rev. Lett., {\bf54}, 1465 (1985).
   
\bibitem{Kirone1}
K. Mallick, P. Marcq, Euro. Phys. J. B, {\bf38}, 99 (2004).

\bibitem{Sebetmoi}
F. P\'etr\'elis and S. Auma\^{\i}tre, 
Eur. Phys. J. B {\bf 34}, 281-284 (2003).
    
\bibitem{VanKampen2} 
N.G. van Kampen, 
{\it Stochastic Process in Physics Chemistry} , North-Holland, Amsterdam, 1992.
\bibitem{bruitdicho}
A. Teubel, U. Behn and A. K\"uhnel, Zeitschrift f\"ur Physik B-Condensed Matter {\bf 71}, 393-402 (1988). I. Bena, C. Van Den Broeck, R. Kawai and K. Lindenberg, Phys. Rev. E {\bf 66}, 045603(R) (2002). We are indebted to C. Van den Broeck for suggesting us the calculation presented in part 2.3.

   
\bibitem{Sawford}
B.L. Sawford, Phys. Fluids A, {\bf 3}, 1577 (1991).

\bibitem{correc}
This expression corrects a misprint for  $M$ defined in \cite{nous}.
     
\bibitem{Lucke2}
M. L\"ucke, {\it Noise in nonlinear dynamical systems}, Vol 2,
Ed. F. Moss \& P.V.E. McClintock, Cambridge 
University Press, 1989.



\bibitem{Heagy} 
J.F. Heagy, N. Platt, S.M. Hammel, Phys. Rev. E, {\bf49}, 
1140 (1994).
   
\bibitem{Cenys}
A. \v{C}enys, A.N. Anagnopoulos, G.L. Bleris, Phys Lett. A,
{\bf224},346 (1997).
   
\bibitem{Lorenz} 
E. Lorenz, Journal of the Atmospheric Sciences, {\bf 20}, 
244 (1963). 

 
\end{thebibliography}
\end{document}